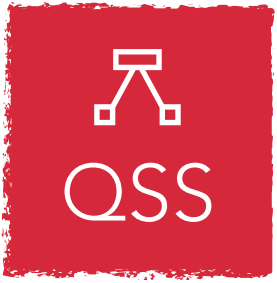

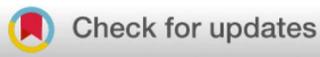

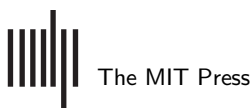

RESEARCH ARTICLE

# Scope and limitations of library metrics for the assessment of ebook usage: COUNTER R5 and link resolver

Mercedes Echeverria[1] 🄳 and Yacelli Bustamante[2] 🄳

[1]Library of the Universidad Autónoma de Madrid, Madrid, Spain
[2]Bioanalysis School, Central University of Venezuela, Caracas, Venezuela

**Keywords:** COUNTER R5, discovery tools, electronic book usage, electronic resource platforms, library metrics, link resolver

## ABSTRACT

Data are at the heart of electronic resource management in academic libraries. Assessing the usage data of electronic resources has become a prevalent approach to demonstrate the value of digital collections, justify library expenditures, and gain insights into how users interact with library materials. This study analyzes the usage statistics of electronic books (ebooks) generated locally by the OpenURL link resolver in an academic library, and statistics collected by platform vendors based on Release 5 of the Counting Online Usage of Networked Electronic Resource (COUNTER R5). Three content provider platforms (Cambridge Core, EBSCOhost, and ScienceDirect) were analyzed as data sources. The COUNTER and link resolver statistics were examined to determine the degree of association between these two metrics. The Spearman correlation coefficient was moderate ($r_s > 0.561$ and $< 0.678$) and statistically significant ($p < .01$). This suggests that these metrics capture different aspects of the usage of ebooks in different contexts. Other factors, such as the types of access to electronic resources and the units of content delivered, were also examined. The study concludes with a discussion regarding the scope and limitations of link resolver and COUNTER R5 as library metrics for measuring the usage of ebooks.

## 1. INTRODUCTION

Academic libraries collect and exploit statistical data for a variety of purposes, including collection management, decision-making, identification of trends, and transparency. Furthermore, in the current landscape, libraries have to deal with the management of a large amount of data generated from multiple websites and platforms (Stuart, 2014), a diversity of metrics and indicators (Mangas-Vega, Gómez-Díaz, & Cordón-García, 2016), and limitations of longitudinal data for tracking trends (Bertot, 2001). Within this framework, measuring the usage of library resources depends on reliable and consistent data. To ensure data consistency for electronic resources, libraries rely on two accepted standards: COUNTER and OpenURL link resolver.

This study examines the usage of ebooks from three content providers' platforms (Cambridge Core, EBSCOhost, and ScienceDirect). This is a case study that utilizes data collected from the Library of Autonomous University of Madrid in Spain, covering the years 2019 to 2021. Data were collected from two sources: the OpenURL link resolver, and the COUNTER R5 reports, obtained via the SUSHI (Standardized Usage Statistics Harvesting Initiative) protocol.

Measuring the usage of electronic resources is in the hands of the platform vendors and beyond the control of libraries (Dulaney & Fernandez, 2019). Therefore, it is important for libraries to collect their own usage data for electronic resources to have a basis for comparison with COUNTER metrics, and to obtain approximate information regarding their electronic resource usage, especially in cases where COUNTER statistics are not available.

The aim of this study is to analyze ebook usage within an academic library context. To address this issue two statistical sources will be analyzed: the usage statistics generated by the Alma link resolver (ExLibris, 2023b) in an academic library setting, and the usage statistics provided by content providers compliant with COUNTER R5 standards.

## 2. LITERATURE REVIEW

Libraries have a strategic interest in enabling users to discover electronic resources and in using the usage data to inform decision-making and improve library management. Consequently, libraries need detailed information on the usage of electronic resources acquired, which are represented in their discovery tool.

This review focused on studies that analyze electronic resource usage in academic libraries using three sources: COUNTER statistics provided by platform vendors, statistics generated by the library's OpenURL link resolver, and a hybrid approach that utilizes both local statistics from the libraries and statistics from platform vendors.

### 2.1. COUNTER

COUNTER is a nonprofit organization supported by a global community of library, publisher, and vendor members, who contribute to the development of the *COUNTER Code of Practice* through working groups and outreach. Since its inception in 2002, COUNTER's primary function has been to provide a code of practice that ensures consistency and comparability of usage statistics for electronic resources. The current *COUNTER Code of Practice Release 5.0.3* (COUNTER, 2023)[1] consists of a set of technical specifications aiming to standardize usage reports across publishers and enable the calculation of Cost-Per-Use (CPU) for ebooks and ejournals (Bull & Beh, 2018; Emery, Estelle, & Adams, 2021; Imre & Estelle, 2021).

In terms of standardization, recent studies have described the new metrics introduced by COUNTER R5 (Bull & Beh, 2018; Getsay & Chen-Gaffey, 2021; Hoeppner, Lendi, & Junge, 2019). Release 5 offers comprehensive and customizable reports that provide detailed information on usage activities, ensuring consistency and comparability of the data. It introduces the Title_Master_Report (TR), which allows for analyzing usage activities performed with ebooks and ejournals at the title level. In particular, the Title_Master_Report (TR) for ebooks allows librarians to assess the impact of ebook titles. Despite the potential usefulness of COUNTER R5 for analyzing ebook usage, there has been limited research literature on this subject. Only a few publications (Chen-Gaffey & Getsay, 2022; Favre, 2020; Jacobs & Hellman, 2023) have examined the metrics related to ebook usage in recent years.

---

[1] The study was conducted using Release 5.0.2 (2021). The changes introduced in Release 5.0.3 (2023) have not affected the metric types used in this study. The changes in Release 5.0.3 are not substantive compared with Release 5.0.2. They consist of corrections and clarifications.

Studies on COUNTER statistics have revealed the potential effect of the platform design on usage counts. Bucknell's study (2012) demonstrated that the interface design can lead to inflated usage statistics. He stated that "Platforms make the HTML full-text the landing page, where the users entitled to view it, so users who want the PDF are forced to record an HTML download en route" (Bucknell, 2012, p. 194). Bergstrom, Uhrig, and Antelman (2018) conducted an analysis of log files, revealing that about 21% of nonbulk downloads constituted PDF files that were downloaded within 5 minutes of a previous HTML download of the same article at the same IP address. Kohn (2018) carried out research on how to calculate an adjustment factor to correct for double counting of HTML and PDF article views.

In response to these biases, Release 5 introduced new metrics to manage the "platform effect" (Imre & Estelle, 2021) and facilitate the comparison of the same units of analysis across platforms (Dulaney & Fernandez, 2019; Emery et al., 2021).

On the other hand, in terms of calculating the CPU, COUNTER R5 lacks coverage for titles with zero usage. This identification is not included in the COUNTER reports because many platforms operate with different systems for authentication and for "subscribed content." Thus, COUNTER reports do not include all zero titles, whether or not the institution subscribed to them. Several authors (Bull & Beh, 2018; Hoeppner et al., 2019; Tingle & Teeter, 2018) have highlighted this issue, emphasizing the importance of zero usage statistics for librarians in CPU calculations, collection value assessment, and potentially revealing the titles that have been purchased but have not been made accessible (Conyers, Lambert et al., 2017).

### 2.2. Link Resolver

The primary goal of link resolver is to provide a seamless and efficient way for users to access the full text of electronic resources identified through a citation or reference. Link resolver is a technology designed to facilitate user access to the appropriate copy of a cited item using the OpenURL protocol. The OpenURL carries the specific metadata of an item to the library's link resolver, which then compares them with the holdings in the library's collection and presents available options on a results page.

In addition to facilitating user access to electronic resources, link resolver logs also provide insights into usage statistics and information about the pathways that users take access to electronic resources (Orcutt, 2010). This functionality allows librarians to track user activity across the discovery tool, repositories, and bibliographic citation software (Chisare, Fagan et al., 2017).

In this scenario, link resolver presents advantages as a valuable source of analytical data for assessing electronic resource usage. Publishers and vendors only provide data on access made exclusively by authenticated users within an institution. Therefore, they are unable to attribute access to an institution when usage is made by users outside the university's IP range or disconnected from its authentication system. In contrast, link resolver provides data on access to electronic resources, including the number of clicks initiated by both authenticated and unauthenticated users, shedding light on how users discover or attempt to access these resources. For instance, requests made through the library's discovery tool could be mentioned in this respect.

However, it is important to note that link reports do not reveal whether a user was able to successfully retrieve the desired article or download the full text (Bealer, 2015; De Groote, Blecic, & Martin, 2013; Jabaily, 2020). In fact, libraries do not collect download statistics directly. Instead, the statistics for full-text downloads are provided to libraries by publishers and vendors (Bergstrom et al., 2018).

In the early stages, link resolver reports for counting local electronic resource usage were often considered imprecise and unreliable. Over the years, librarians have raised concerns with vendors regarding the quality of metadata, accuracy of knowledge base holdings, and link syntax generated by this software.

Wakimoto, Walker, and Dabbour (2006) conducted keyword searches in several databases to assess the effectiveness of the SFX link resolver[2]. The study found that around 20% of the full-text options provided by SFX were inaccurate. These inaccuracies were categorized into false positives and false negatives. False positives occurred when SFX incorrectly displayed the availability of a full-text article, and false negatives happened when SFX failed to display the availability of a full-text article. Further analysis indicated that most false negatives were caused by incorrect holding information provided by database providers or by providers who failed to load specific articles. Conversely, most false positives were attributed to OpenURLs generated incorrectly by the source databases, which sent inaccurate information to SFX.

Chandler (2009) analyzed the OpenURLs in the *L'Année philologique* database to evaluate the quality of OpenURL links based on the presence or absence of specific metadata. The study aimed to serve as a proof of concept for the overall quality of OpenURL links. The analysis revealed a range of metadata problems in the OpenURL links, including malformed dates, volume and issue numbers combined into a single field, reliance on the "pages" element instead of the "start page element," and a lack of identifiers, among other concerns. These issues could effectively impact the accuracy and reliability of the OpenURL links, increasing the likelihood of incorrect and broken links.

Trainor and Price (2010) analyzed the accuracy rates and the causes of failure in link resolver in a random sample of OpenURLs. They found an overall precision rate of 71% for the citations tested. The study revealed that error causes were evenly distributed across the OpenURL resolver chain, involving incorrect data from the source, translation errors from the link resolver, and problems with the target understanding or acceptance of the OpenURL requests.

In summary, the authors indicated that the quality of OpenURL metadata is critical for the linking model. Although some studies recommended collaborating with content providers to improve access to electronic resources, Chandler's approach went beyond this. The initial results of the Cornell study convinced Chandler to submit a proposal to the National Information Standards Organization (NISO). This initiative gave way to the Improving OpenURLs Through Analytics (IOTA) Working Group in early 2010.

Although challenges still exist (Hills, 2020; Rathemacher, Ragucci, & Doellinger, 2022), significant advances have been made in the exchange of metadata between content providers and libraries' knowledge bases. In this regard, the recommendation from NISO's Knowledge Base and Related Tools (KBART) has played a crucial role in establishing a framework that facilitates the transfer of holdings metadata from content providers to knowledge base suppliers and libraries. KBART was initiated by UKSG and now is a NISO Recommended Practice. In 2014, Pesch (2014) outlined the recommendations for KBART Phase I and Phase II, which aimed to improve OpenURL linking. KBART Phase III was initiated in 2019 and has introduced substantial additions and new areas of work. Its goals include improving support for additional content types and global content, clarifying endorsement procedures, providing holdings at more granular levels, and offering additional file formats such as JSON and XML (NISO KBART Standing Committee, 2021).

---

[2] https://exlibrisgroup.com/products/primo-discovery-service/sfx-link-resolver/.

KBART has been widely adopted by discovery systems to ensure the accuracy of institutional holdings and facilitate the transfer of holding data between content providers and knowledge bases, and thus reduce the effort required to maintain electronic resources in library systems.

The fact that libraries and discovery service providers have continued to support link resolver software over the years highlights the importance of this tool. This observation would suggest that libraries need to collect specific usage information from the discovery tools because this information is essential for contextualizing electronic resource usage within the library environment.

### 2.3. Hybrid Statistics

Given the number of factors that may influence the usage of electronic resources, some libraries have resorted to collecting data locally to gain a more comprehensive understanding. This section examines the cases where link resolver and COUNTER reports have been compared for evaluating the usage of electronic resources.

Bordeaux and Kramer made a presentation on library usage statistics at the North American Serial Interest Group in 2004. They discussed several options for gathering and compiling usage data. Bordeaux, from Binghamton University, tracked the statistics of 93 databases using COUNTER and link resolver reports. She found that the "information that vendor-supplied statistics will not provide libraries" (Bordeaux, Kraemer, & Sullenger, 2005, p. 298). Kraemer, from the University of Wisconsin, compared the usage data from COUNTER reports to the click-through counts obtained from OpenURLs. He concluded that "[click-through counts] are great for estimating use data where the vendor will not supply statistics" (Bordeaux et al., 2005, p. 296).

De Groote et al. (2013) assessed electronic journal usage through COUNTER and link resolver reports. They found a strong positive correlation between the two metrics (Spearman's $r_s = 0.843$). Based on this result, the authors concluded that "[this result] allows libraries to substitute link-resolver data for vendor data in most circumstances" (De Groote et al., 2013, p. 110).

Greenberg and Bar-Ilan (2017) conducted an analysis of library users' information retrieval behavior using log files, reports, and publishers' counts at the University of Haifa. The study revealed that the number of full-text downloads based on COUNTER requests was 3.1 times higher than the number of downloads requested through the discovery tool, as documented by the library's OpenURL link resolver service.

Overall, the studies reviewed reflect that libraries heavily rely on COUNTER data. Additionally, they reveal a significant difference between the number of requests from the library's discovery tool and the requests reported by content providers. This variation highlights the need for libraries to obtain additional insights into how their resources are accessed and utilized, both within and outside the library environment.

### 2.4. Aims and Objectives

Despite the potential usefulness that the library metrics approach may have for gaining a comprehensive view of electronic resource usage, no previous studies have integrated COUNTER and link resolver metrics to analyze ebook usage. This study takes a statistical approach to analyzing ebook usage, utilizing the content provider's COUNTER R5 statistics and the statistics of the link resolver clicked requests obtained from the library's discovery tool.

We aim to determine the degree of correlation between COUNTER and link resolver metrics, and to identify the factors influencing the discoverability and usage of ebooks. To achieve this, we outline the following objectives:

1. Determine the degree of correlation between COUNTER and link resolver statistics.
2. Evaluate the factors that may influence the usage of ebooks.

## 3. METHODS AND DATA

The research was conducted in the Library of Autonomous University of Madrid (UAM), a medium-sized university in Spain. According to 2021 data, the UAM had an enrollment of more than 30,000 students, comprising 21,000 undergraduates and 10,000 postgraduates, as well as a faculty of 3,300 members, and an additional staff of 1,200 members. In 2021, the UAM Library provided access to nearly 700,000 ebook portfolios from paid collections through its discovery tool. As a member of the Madroño Consortium, the UAM Library is part of a network that includes five other university libraries in Madrid. In spring 2018, the Library implemented ExLibris's Alma-Primo discovery tool, utilizing Primo Back Office, in addition to Alma link resolver and Alma Analytics.

The approach for analyzing ebook usage includes the following sections:

- data sources: selection of content provider platforms;
- data collection: COUNTER and link resolver reports; and
- data analysis: statistical methods used to measure ebook usage.

### 3.1. Data Sources

The starting point of this study was the selection of the ebook platforms that provided research content along with usage metrics. For selection purposes, the ebook platforms were chosen based on the following considerations: electronic collection type, access functionalities, and the unit of content delivered, as detailed in Table 1.

### 3.2. Data Collection

The usage data of the content providers' platforms was collected in January 2022. This study examines ebook usage from January 2019 to December 2021. This section is divided into three parts: sample size, analysis of the COUNTER reports, and analysis of the link resolver reports.

#### 3.2.1. Sample size

Table 2 summarizes the descriptive statistics of the sample size used in this study. It includes the number of ebook portfolios of each ebook collection, obtained from E-Inventory, a subject area of Alma Analytics. In this context, a portfolio is defined as "an electronic resource (inventory) that maintains the specific coverage (local or global), services, and link information relevant for a particular electronic title that may also be part of an electronic collection" (ExLibris, 2023a). An electronic portfolio can contain an ejournal, an ebook, a streaming video, or other types of materials. Furthermore, Table 2 details the number of ebook titles analyzed in the sample of this study, obtained from the COUNTER reports via SUSHI, and the number of ebook titles analyzed in this sample of the study obtained from the Link Resolver reports.

**Table 1.** Ebook platforms and selection criteria

| Platforms | Collection type | Access functionalities | Unit of content delivered |
|---|---|---|---|
| Cambridge Core | Selective package of Cambridge University Press | No restrictions on accessing the table of contents of licensed or nonlicensed ebooks | Chapter level and the entire ebook |
| EBSCOhost | Aggregator | Restrictions on accessing the table of contents of licensed ebooks | Chapter level |
| ScienceDirect | Selective package of Elsevier | No restrictions on accessing the table of contents of licensed or nonlicensed ebooks | Chapter level and the entire ebook |

**Table 2.** Ebook portfolios and titles in the sample size

| | Platforms | | | | | | | | |
|---|---|---|---|---|---|---|---|---|---|
| | **Cambridge Core** | | | **EBSCOhost** | | | **ScienceDirect** | | |
| **Portfolios and titles** | **2019** | **2020** | **2021** | **2019** | **2020** | **2021** | **2019** | **2020** | **2021** |
| Number of ebook portfolios (E-Inventory) | 452 | 507 | 522 | 8,674 | 8,834 | 230,536 | 2,263 | 2,471 | 2,680 |
| Number of ebook titles (COUNTER_SUSHI API) | 1,404 | 1,340 | 1,515 | 1,091 | 1,549 | 5,647 | 3,927 | 4,554 | 4,377 |
| Number of ebook titles (link resolver) | 56 | 87 | 82 | 1,023 | 2,243 | 5,600 | 238 | 230 | 261 |

It is worth mentioning that the EBSCOhost Ebook Academic Collection-World Wide was expanded on September 7, 2021 with the activation of 221,893 portfolios.

### 3.2.2. COUNTER reports

The COUNTER usage data were automatically harvested via SUSHI, a standard (ANSI/NISO Z39.93-2014) developed by the National Information Standards Organization (NISO, 2014) that enables the automated harvesting of COUNTER reports. The platforms, Cambridge Core, EBSCOhost, and ScienceDirect, support the automatic harvesting of COUNTER reports via the COUNTER_SUSHI API, which uses a RESTful interface to access and retrieve COUNTER reports in JSON (JavaScript Object Notation) format. The SUSHI API is provided by each content provider, allowing Alma Analytics to automatically retrieve COUNTER statistics reports on library resource usage.

The COUNTER reports contain the usage data for both licensed and gold open access books. Furthermore, the TR_B2 No_License metric provides data on the access to nonlicensed ebooks by authenticated users on the Cambridge Core and ScienceDirect platforms. The TR_B2 Limit_Exceeded metric includes turn-away data for books where users were denied access due to simultaneous usage (concurrency) or licenses exceeded.

The COUNTER reports were generated in Alma Analytics according to the following criteria: material type indicators, normalized title, ISBN prompts, access type prompts, and years of usage.

Table 3 displays the COUNTER reports harvested through the SUSHI API, broken down by platform name, year of usage, and applicable COUNTER metrics for ebooks. The data are detailed separately for the years 2019–2021. The COUNTER metrics are defined below for a better interpretation of data according to the *COUNTER Code of Practice* (COUNTER, 2021).

- TR_B2 Limit_Exceeded. A Metric_Type. A user is denied access to a content item because the simultaneous-user limit for their institution's license would be exceeded.
- TR_B2 No_License. A Metric_Type. A user is denied access to a content item because the user or the user's institution does not have access rights under an agreement with the vendor.

- TR_B3 Total_Item_Investigations. A Metric_Type that represents the number of times users accessed the content (e.g., a full text) of an item, or information describing that item (e.g., an abstract).
- TR_B3 Total_Item_Requests. A Metric_Type that represents the number of times users requested the full content (e.g., a full text) of an item. Requests may take the form of viewing, downloading, emailing, or printing content, provided such actions can be tracked by the content provider.
- TR_B3 Unique_Item_Investigations. A Metric_Type that represents the number of unique content items investigated in a user session. Examples of content items are articles, book chapters, and multimedia files.
- TR_B3 Unique_Item_Requests. A Metric_Type that represents the number of unique content items requested in a user session. Examples of content items are articles, book chapters, and multimedia files.
- TR_B3 Unique_Title_Investigations. A Metric_Type that represents the number of unique titles investigated in a user session. This Metric_Type is only applicable for Data_Type Book.
- TR_B3 Unique_Title_Requests. A Metric_Type that represents the number of unique titles requested in a user session. This Metric_Type is only applicable for Data_Type Book.

The term *request* is specially related to viewing or downloading the full-text of an entire book or a book chapter, and the term *investigation* encompasses any action performed by a user in relation to a content item or title, including viewing or downloading the full text. Therefore, investigation reports may include both requests and investigations, whereas request reports are specifically related to the viewing or downloading of the content of the item itself (e.g., the full text of an ebook or a book chapter).

Regarding Table 3, it should be noted that the data for ScienceDirect in 2019 appears to differ from the values reported for 2020 and 2021 in several metrics: TR_B2 No_License, TR_B3 Total_Item_Investigations, and TR_B3 Unique_Item_Investigations, and TR_B3 Unique_Title_Investigations. The reason for these discrepancies is not clear and could be related to various factors, such as the transition from COUNTER 4 to COUNTER 5 metrics, the impact of COVID-19 lockdowns, or other factors that have not been identified.

**Table 3.** COUNTER R5 metrics, 2019–2021. Data harvested using SUSHI

| | Platforms | | | | | | | | |
| | Cambridge Core | | | EBSCOhost | | | ScienceDirect | | |
| COUNTER R5 metrics | 2019 | 2020 | 2021 | 2019 | 2020 | 2021 | 2019 | 2020 | 2021 |
|---|---|---|---|---|---|---|---|---|---|
| TR_B2 Limit_Exceeded | 0 | 0 | 0 | 5 | 24 | 78 | 0 | 0 | 0 |
| TR_B2 No_License | 4,152 | 3,443 | 3,669 | 0 | 0 | 0 | 34,970 | 1,962 | 16,583 |
| TR_B3 Total_Item_Investigations | 6,860 | 6,679 | 5,887 | 5,195 | 4,124 | 20,434 | 72,816 | 33,286 | 30,418 |
| TR_B3 Total_Item_Requests | 3,101 | 2,975 | 2,319 | 1,468 | 2,835 | 10,019 | 27,357 | 18,471 | 9,978 |
| TR_B3 Unique_Item_Investigations | 5,833 | 5,530 | 4,686 | 4,156 | 2,973 | 10,885 | 61,201 | 27,896 | 25,114 |
| TR_B3 Unique_Item_Requests | 2,763 | 2,510 | 1,927 | 1,310 | 2,065 | 6,900 | 23,766 | 15,601 | 8,381 |
| TR_B3 Unique_Title_Investigations | 2,623 | 2,668 | 2,320 | 3,226 | 2,630 | 10,537 | 30,799 | 11,081 | 13,826 |
| TR_B3 Unique_Title_Requests | 338 | 389 | 300 | 719 | 1,814 | 6,688 | 4,671 | 3,792 | 1,830 |

### 3.2.3. Link resolver

Link resolver is a powerful tool used by libraries to provide seamless access to electronic resources. It enables users to check the availability of an electronic resource directly from a citation while they are conducting their search. Link resolver uses the metadata embedded in the OpenURL of the requested item to search in the institution's knowledge base (e.g., Alma), and determine whether the institution has access to the resource. If the appropriate copy is found, link resolver will then send this data to the content provider, which will respond to the request by delivering the desired content to the user.

The link resolver reports were extracted from "Link resolver usage," a subject area of Alma Analytics. The list of ebook titles licensed by each content provider was obtained using "E-Inventory," another subject area of Alma Analytics. These titles were subsequently uploaded to the "Link resolver usage." This process was necessary because "Link resolver usage" did not provide complete data by platform source.

The reports were parameterized according to the following criteria: material type indicators, normalized title, ISBN prompts, and requests from the years 2019 to 2021. In Alma Link Resolver reports, the "source type," which is the field that allows libraries to identify where users are coming from, was set to Primo/Primo Central.

The link resolver reports required a data-cleaning process to address duplicate requests produced within a time frame of less than 1 minute. Additional actions involved removing clicked requests with a value "0," indicating that no services were accessed. Finally, the OpenURLs of each request were examined to verify if the OpenURLs pointed to the specific platform being analyzed.

Table 4 details the link resolver reports for ebooks broken down by platform name, the number of titles, and all applicable link resolver metrics. The data is detailed separately for the years 2019–2021.

It is important to note that link resolver reports for Cambridge Core and ScienceDirect include data on both licensed and nonlicensed ebooks, accessed by both authenticated and nonauthenticated users through the discovery tool. In contrast, link resolver reports for EBSCOhost provide data on accesses made to licensed books. However, the EBSCOhost reports also capture the data on attempts made by nonauthenticated users through the discovery tool, although those attempts were unsuccessful in accessing ebooks.

**Table 4.** Link resolver metrics, 2019–2021

| | Platforms | | | | | | | | |
| | Cambridge Core | | | EBSCOhost | | | ScienceDirect | | |
| Link resolver metrics | 2019 | 2020 | 2021 | 2019 | 2020 | 2021 | 2019 | 2020 | 2021 |
| --- | --- | --- | --- | --- | --- | --- | --- | --- | --- |
| Number of requests | 118 | 226 | 158 | 1,607 | 3,647 | 8,991 | 703 | 868 | 1,056 |
| Number of clicked requests | 117 | 224 | 157 | 1,607 | 3,647 | 8,991 | 699 | 851 | 1,036 |
| Number of requests without services | 0 | 0 | 0 | 0 | 0 | 0 | 0 | 0 | 0 |
| Number of services (total) | 119 | 281 | 227 | 2,224 | 5,522 | 12,137 | 757 | 966 | 1,303 |
| Number of electronic services | 119 | 281 | 227 | 2,244 | 5,522 | 12,135 | 757 | 966 | 1,303 |
| Number of clicked services (total) | 117 | 226 | 163 | 1,621 | 3,897 | 9,422 | 706 | 866 | 1,056 |
| Number of electronic clicked services | 117 | 226 | 163 | 1,621 | 3,897 | 9,422 | 706 | 866 | 1,056 |

### 3.3. Data Analysis

Nonparametric statistical tests were utilized because the data did not follow a normal distribution. To assess the degree of association between COUNTER and link resolver metrics the Spearman correlation test was used. To demonstrate differences between platforms, the Kruskal-Wallis test and comparisons with Bonferroni adjustments were used. The Mann-Whitney U test was used to determine whether two groups were significantly different from each other in the variable of interest. Descriptive statistics such as the median and interquartile range (IQR) were utilized. Statistical significance was determined with a probability of error of less than 5% ($p < .05$).

The COUNTER and link resolver data were exported from Alma Analytics to Microsoft Excel. The statistical analyses were performed using IBM SPSS software (Version 28.0.1.1).

## 4. RESULTS

The specific outcomes of this study are presented below.

### 4.1. Correlation Between COUNTER and Link Resolver Metrics

The Spearman correlation test was used to measure the degree of association between COUNTER and link resolver metrics. The variables used in the analysis were the COUNTER metric Unique_Title_Investigations, and the link resolver metric Clicked Requests.

The Clicked Requests metric was used because it indicates the number of OpenURLs generated to access ebooks, regardless of whether users accessed the full text or not. It is worth noting that the actions performed by users through link resolver are categorized as investigations, according to COUNTER R5 (Mellins-Cohen, 2018).

The Unique_Title_Investigations metric was used for analysis as it enables the comparison of books, regardless of their delivery mechanism, either as a whole book or as individual chapters.

The analysis consisted of identifying matching requests between COUNTER and link resolver reports according to the following parameters: the same usage year, the same title, and ISBN.

Spearman correlation tests were calculated individually for each specific platform to determine the degree of association between COUNTER and link resolver metrics. These correlations were all positive and statistically significant at the $p = .01$ level (Tables 5, 6, and 7). The highest positive correlation was found for Cambridge Core ($r_s = 0.678$), followed by EBSCOhost ($r_s = 0.656$) and ScienceDirect ($r_s = 0.561$).

Furthermore, the analysis revealed small differences in the correlation coefficients between the metrics Unique_Title_Investigations and Clicked Requests, and Unique_Title_Requests and Clicked Requests. In addition, the platforms maintain a consistent order in the results of the analyzed metrics across all platforms.

Specifically, the correlation coefficients for Unique_Title_Investigations and Clicked Requests were as follows: Cambridge Core ($r_s = 0.678$), EBSCOhost ($r_s = 0.656$), and Science-Direct ($r_s = 0.561$). On the other hand, the correlation coefficients for Unique_Title_Requests and Clicked Requests were as follows: Cambridge Core ($r_s = 0.602$), EBSCOhost ($r_s = 0.553$), and ScienceDirect ($r_s = 0.519$) (Tables 5, 6, and 7).

Similarly, the differences found in the correlation coefficients for the following metrics were small: Unique_Item_Investigations and Clicked Requests, and Unique_Item_Requests and Clicked Requests. In addition, a consistent platform order based on the correlation coefficients was observed across all platforms.

**Table 5.** Cambridge Core. Spearman correlation between COUNTER and link resolver metrics

| Spearman's rho CAMBRIDGE | TR_B3 – Total Item Investigations | TR_B3 – Total Item Requests | TR_B3 – Unique Item Investigations | TR_B3 – Unique Item Requests | TR_B3 – Unique Title Investigations | TR_B3 – Unique Title Requests | Number of Requests | Number of Clicked Requests | Number of Services (total) | Number of Electronic Services | Number of Clicked Services (total) | Number of Electronic Clicked Services |
|---|---|---|---|---|---|---|---|---|---|---|---|---|
| TR_B3 – Total Item Investigations | 1.000 | 0.953** | 0.982** | 0.944** | 0.640** | 0.815** | 0.566** | 0.566** | 0.513** | 0.513** | 0.566** | 0.566** |
| TR_B3 – Total Item Requests | 0.953** | 1.000 | 0.941** | 0.989** | 0.458** | 0.775** | 0.445** | 0.445** | 0.396** | 0.396** | 0.447** | 0.447** |
| TR_B3 – Unique Item Investigations | 0.982** | 0.941** | 1.000 | 0.952** | 0.650** | 0.793** | 0.541** | 0.541** | 0.498** | 0.498** | 0.540** | 0.540** |
| TR_B3 – Unique Item Requests | 0.944** | 0.989** | 0.952** | 1.000 | 0.463** | 0.779** | 0.447** | 0.447** | 0.398** | 0.398** | 0.447** | 0.447** |
| TR_B3 – Unique Title Investigations | 0.640** | 0.458** | 0.650** | 0.463** | 1.000 | 0.717** | 0.678** | 0.678** | 0.655** | 0.655** | 0.675** | 0.675** |
| TR_B3 – Unique Title Requests | 0.815** | 0.775** | 0.793** | 0.779** | 0.717** | 1.000 | 0.602** | 0.602** | 0.534** | 0.534** | 0.598** | 0.598** |
| Number of Requests | 0.566** | 0.445** | 0.541** | 0.447** | 0.678** | 0.602** | 1.000 | 1.000** | 0.920** | 0.920** | 0.999** | 0.999** |
| Number of Clicked Requests | 0.566** | 0.445** | 0.541** | 0.447** | 0.678** | 0.602** | 1.000** | 1.000 | 0.920** | 0.920** | 0.999** | 0.999** |
| Number of Services (total) | 0.513** | 0.396** | 0.498** | 0.398** | 0.655** | 0.534** | 0.920** | 0.920** | 1.000 | 1.000** | 0.923** | 0.923** |
| Number of Electronic Services | 0.513** | 0.396** | 0.498** | 0.398** | 0.655** | 0.534** | 0.920** | 0.920** | 1.000** | 1.000 | 0.923** | 0.923** |
| Number of Clicked Services (total) | 0.566** | 0.447** | 0.540** | 0.447** | 0.675** | 0.598** | 0.999** | 0.999** | 0.923** | 0.923** | 1.000 | 1.000** |
| Number of Electronic Clicked Services | 0.566** | 0.447** | 0.540** | 0.447** | 0.675** | 0.598** | 0.999** | 0.999** | 0.923** | 0.923** | 1.000** | 1.000 |

** Correlation is at $p = .01$.

**Table 6.** EBSCOhost. Spearman correlation between COUNTER and link resolver metrics

| Spearman's rho EBSCOhost | TR_B3 – Total Item Investigations | TR_B3 – Total Item Requests | TR_B3 – Unique Item Investigations | TR_B3 – Unique Item Requests | TR_B3 – Unique Title Investigations | TR_B3 – Unique Title Requests | Number of Requests | Number of Clicked Requests | Number of Services (total) | Number of Electronic Services | Number of Clicked Services (total) | Number of Electronic Clicked Services |
|---|---|---|---|---|---|---|---|---|---|---|---|---|
| TR_B3 - Total Item Investigations | 1.000 | 0.882** | 0.789** | 0.802** | 0.771** | 0.791** | 0.538** | 0.538** | 0.473** | 0.472** | 0.520** | 0.520** |
| TR_B3 – Total Item Requests | 0.882** | 1.000 | 0.667** | 0.888** | 0.643** | 0.874** | 0.475** | 0.475** | 0.421** | 0.420** | 0.461** | 0.461** |
| TR_B3 – Unique Item Investigations | 0.789** | 0.667** | 1.000 | 0.775** | 0.974** | 0.754** | 0.641** | 0.641** | 0.568** | 0.569** | 0.623** | 0.623** |
| TR_B3 – Unique Item Requests | 0.802** | 0.888** | 0.775** | 1.000 | 0.747** | 0.983** | 0.541** | 0.541** | 0.482** | 0.482** | 0.524** | 0.524** |
| TR_B3 – Unique Title Investigations | 0.771** | 0.643** | 0.974** | 0.747** | 1.000 | 0.766** | 0.656** | 0.656** | 0.584** | 0.585** | 0.638** | 0.638** |
| TR_B3 – Unique Title Requests | 0.791** | 0.874** | 0.754** | 0.983** | 0.766** | 1.000 | 0.553** | 0.553** | 0.495** | 0.496** | 0.536** | 0.536** |
| Number of Requests | 0.538** | 0.475** | 0.641** | 0.541** | 0.656** | 0.553** | 1.000 | 1.000** | 0.902** | 0.902** | 0.975** | 0.975** |
| Number of Clicked Requests | 0.538** | 0.475** | 0.641** | 0.541** | 0.656** | 0.553** | 1.000** | 1.000 | 0.902** | 0.902** | 0.975** | 0.975** |
| Number of Services (total) | 0.473** | 0.421** | 0.568** | 0.482** | 0.584** | 0.495** | 0.902** | 0.902 | 1.000 | 1.000** | 0.929** | 0.929** |
| Number of Electronic Services | 0.472** | 0.420** | 0.569** | 0.482** | 0.585** | 0.496** | 0.902** | 0.902** | 1.000** | 1.000 | 0.929** | 0.929** |
| Number of Clicked Services (total) | 0.520** | 0.461** | 0.623** | 0.524** | 0.638** | 0.536** | 0.975** | 0.975** | 0.929** | 0.929** | 1.000 | 1.000** |
| Number of Electronic Clicked Services | 0.520** | 0.461** | 0.623** | 0.524** | 0.638** | 0.536** | 0.975** | 0.975** | 0.929** | 0.929** | 1.000** | 1.000 |

** Correlation is at $p = .01$.

**Table 7.** ScienceDirect. Spearman correlation between COUNTER and link resolver metrics

| Spearman's rho ScienceDirect | TR_B3 – Total Item Investigations | TR_B3 – Total Item Requests | TR_B3 – Unique Item Investigations | TR_B3 – Unique Item Requests | TR_B3 – Unique Title Investigations | TR_B3 – Unique Title Requests | Number of Requests | Number of Clicked Requests | Number of Services (total) | Number of Electronic Services | Number of Clicked Services (total) | Number of Electronic Clicked Services |
|---|---|---|---|---|---|---|---|---|---|---|---|---|
| TR_B3 – Total Item Investigations | 1.000 | 0.980** | 0.633** | 0.975** | 0.727** | 0.699** | 0.463** | 0.465** | 0.452** | 0.456** | 0.446** | 0.446** |
| TR_B3 – Total Item Requests | 0.980** | 1.000 | 0.638** | 0.989** | 0.738** | 0.749** | 0.455** | 0.456** | 0.438** | 0.439** | 0.436** | 0.436** |
| TR_B3 – Unique Item Investigations | 0.633** | 0.638** | 1.000 | 0.617** | 0.710** | 0.685** | 0.447** | 0.445** | 0.407** | 0.409** | 0.428** | 0.428** |
| TR_B3 – Unique Item Requests | 0.975** | 0.989** | 0.617** | 1.000 | 0.695** | 0.709** | 0.432** | 0.433** | 0.420** | 0.421** | 0.416** | 0.416** |
| TR_B3 – Unique Title Investigations | 0.727** | 0.738** | 0.71** | 0.695** | 1.000 | 0.948** | 0.563** | 0.561** | 0.522** | 0.520** | 0.541** | 0.541** |
| TR_B3 – Unique Title Requests | 0.699** | 0.749** | 0.685** | 0.709** | 0.948** | 1.000 | 0.521** | 0.519** | 0.48** | 0.473** | 0.498** | 0.498** |
| Number of Requests | 0.463** | 0.455** | 0.447** | 0.432** | 0.563** | 0.521** | 1.000 | 0.998** | 0.939** | 0.936** | 0.968** | 0.968** |
| Number of Clicked Requests | 0.465** | 0.456** | 0.445** | 0.433** | 0.561** | 0.519** | 0.998** | 1.000 | 0.937** | 0.936** | 0.969** | 0.969** |
| Number of Services (total) | 0.452** | 0.438** | 0.407** | 0.420** | 0.522** | 0.480** | 0.939** | 0.937** | 1.000 | 0.997** | 0.924** | 0.924** |
| Number of Electronic Services | 0.456** | 0.439** | 0.409** | 0.421** | 0.520** | 0.473** | 0.936** | 0.936** | 0.997** | 1.000 | 0.923** | 0.923** |
| Number of Clicked Services (total) | 0.446** | 0.436** | 0.428** | 0.416** | 0.541** | 0.498** | 0.968** | 0.969** | 0.924** | 0.923** | 1.000 | 1.000** |
| Number of Electronic Clicked Services | 0.446** | 0.436** | 0.428** | 0.416** | 0.541** | 0.498** | 0.968** | 0.969** | 0.924** | 0.923** | 1.000** | 1.000 |

** Correlation is at $p = .01$.

Specifically, the correlation coefficients for Unique_Item_Investigations and Clicked Requests were as follows: EBSCOhost ($r_s = 0.641$), Cambridge Core ($r_s = 0.541$) and ScienceDirect ($r_s = 0.445$). On the other hand, the correlation coefficients for Unique_Item_Requests and Clicked Requests were as follows: EBSCOhost ($r_s = 0.541$), Cambridge Core ($r_s = 0.447$) and ScienceDirect ($r_s = 0.433$) (see Tables 5, 6, and 7). For detailed results of the Spearman correlation (Supplementary material: Tables S1, S2, and S3).

The analysis showed that the correlations for Investigations and Clicked Requests were stronger than those for Requests and Clicked Requests. Additionally, the small differences observed in the correlation coefficients between Investigations and Requests suggest a close relationship between these two metrics.

### 4.2. Differences Between Platforms

The aim of this analysis was to determine whether there were differences between platforms.

The Kruskal-Wallis test is a nonparametric test used to compare unpaired groups, when data does not follow a normal distribution. In this study, the Kruskal-Wallis test was used to assess whether there were differences among platforms. The groups analyzed were the three platforms with their respective matched requests between COUNTER and link resolver.

The Kruskal-Wallis test showed a statistically significant difference among the platforms regarding the number of Clicked Requests (Kruskal-Wallis = 267.36, $p = .000$).

After applying the Bonferroni correction for multiple comparisons, pairwise comparisons showed statistically significant differences between Cambridge Core and EBSCOhost ($z = 9.78$, $p = .000$), and ScienceDirect and EBSCOhost ($z = 13.461$, $p = .000$). In contrast, no statistically significant differences were found between Cambridge Core and ScienceDirect ($z = 2.151$, $p = .094$).

These findings suggest that both Cambridge Core and ScienceDirect have similar patterns in terms of ebook accessibility from the discovery tool, as both platforms allow users to access both licensed and nonlicensed ebooks, regardless of whether users are connected to the university's authentication system or not.

### 4.3. Factors That May Influence Ebook Usage

This section examines the factors that can influence the usage of ebooks. The analysis is divided into two parts. The first part investigates the types of ebook access, and the second part is focused on the unit of content delivered, whether as individual chapters or as entire books.

#### 4.3.1. Ebook access types

This section compares two models of accessing ebooks: first, the EBSCOhost platform, which requires users to be connected to the university's authentication system for accessing licensed ebooks, and second, the Cambridge Core and ScienceDirect platforms, which allow users to access the table of contents of both licensed and nonlicensed ebooks from the discovery tool without the need to be connected to the university's authentication system. However, it is worth noting that viewing or downloading of the full text is only possible when users are connected to the university's authentication system.

The Mann-Whitney U test is a nonparametric test used to determine differences between groups with respect to a factor. In this study, the types of ebook access were analyzed. We compared the platforms that require the university's authentication with those that do not require it. The results showed statistically significant differences regarding ebook access types (Mann-Whitney U = 1,811,815, $z = -16.218$, $p < .011$).

The comparison of access types between platforms showed significant differences in the median.

- Platforms requiring university authentication: median = 1.0 (IQR 1.0).
- Platforms not requiring university authentication: median = 2.0 (IQR 3.0).

### 4.3.2. Unit of content delivered

The second factor analyzed was the unit of content delivered, specifically whether it was a chapter or an entire book. It is worth noting that any user activity related to a content item, including downloading or viewing the full text, is considered as "Investigations" according to COUNTER R5.

Theoretical scenarios for analyzing these factors are as follows:

- If a publisher provides ebooks as a single PDF, it is reasonable to expect a correlation between Unique_Title_Investigations and clicked investigations/requests reported by the link resolver.
- If a publisher provides ebooks by chapters, as is the case with EBSCOhost, a significant correlation between the Unique_Item_Investigations metric and clicked investigations/requests reported by the link resolver is expected. It is interesting to note that the first click on a chapter is also counted in Unique_Title_Investigations, although any subsequent clicks in the same book during the same session are not counted (Hendry, 2020). In summary, Unique_Title_Investigations records when a user is interested in and interacts with a book.
- When a content provider uses a hybrid model to deliver ebook content, either as an entire book or by chapters, as is the case with Cambridge Core and ScienceDirect, the correlation between Unique_Title_Investigations and Unique_Item_Investigations with clicked investigations/requests reported by link resolver cannot be estimated because the disaggregated data is unknown. This is because the Unique_Title_Investigations metric counts the first click made at the chapter level, and therefore it is unclear how the book was delivered, whether as a whole book or by chapters.

**Table 8.** Spearman correlation between Clicked Requests reported by link resolver and Unique_Title_Investigations and Unique_Item_Investigations metrics

| | | Correlation of Clicked Requests reported by link resolver and: | | | | | |
| --- | --- | --- | --- | --- | --- | --- | --- |
| | | Unique_Title_Investigations [COUNTER] | | | Unique_Item_Investigations [COUNTER] | | |
| | | | 95% Confidence intervals | | | 95% Confidence intervals | |
| Platforms | N | Spearman's $\rho$ | Lower | Upper | Spearman's $\rho$ | Lower | Upper |
| Cambridge Core | 92 | 0.678** | 0.548 | 0.779 | 0.541** | 0.378 | 0.671 |
| EBSCOhost | 3,754 | 0.656** | 0.631 | 0.680 | 0.641** | 0.615 | 0.668 |
| ScienceDirect | 263 | 0.561** | 0.467 | 0.646 | 0.445** | 0.330 | 0.537 |

** Significant at $p = .01$.

Based on the previous findings, it was observed that Unique_Title_Investigations showed the highest Spearman correlation across all three platforms. Consequently, it is expected that Unique_Title_Investigations will achieve the highest correlation coefficient in whatever comparisons to be made between COUNTER and link resolver.

The Unique_Item_Investigations metric measures ebook usage at the chapter level. As expected, the highest correlation between Unique_Item_Investigations and investigations/requests reported by the link resolver was found for EBSCOhost, which delivers ebooks as chapters (Table 8).

## 5. DISCUSSION

The study found moderate positive correlations that were statistically significant between the COUNTER metrics (Unique_Title_Investigations and Unique_Item_Investigations) with link resolver Clicked Requests for the ebook platforms (Cambridge Core, EBSCOhost, and ScienceDirect). The results indicate variations in ebook usage among these platforms, with Spearman correlation coefficients ranging from $r_s = 0.561$ to $r_s = 0.678$ at $p = .01$. These moderate correlations suggest that COUNTER and link resolver metrics capture different aspects of ebook usage in different contexts.

These results diverge from the findings reported by De Groote et al. (2013) for electronic journal usage, where the Spearman correlation achieved a coefficient of $r_s = 0.843$. This difference in the correlations could suggest that the factors affecting the usage and reporting of electronic journals may be different from those affecting ebooks.

The following aspects were considered to contextualize the results.

### 5.1. Scope of the Library Metrics

At present, COUNTER reports are the sole means available to librarians for obtaining consistent data on the viewing and downloading of full-text content items.

COUNTER Release 5 has introduced new metrics along with a series of elements and attributes to report the usage of ebooks in a more granular manner. Among these, the Unique_Title metric is specifically designed for ebooks and allows librarians to compare usage counts of the same content unit across different platforms. Release 5 also provides detailed information about different types of usage, including "investigations" and "requests," which measure the level of interaction of users with the ebook contents. Additionally, Section_types identify the specific ebook content accessed by users, including "book" and "chapter."

Link resolver information is essential for librarians, as it is the only means through which libraries can obtain the data on the usage of electronic resources from the discovery tool and see the effect of discovery regarding the information supplied by content providers.

Recent studies have examined how users utilize the discovery tool. Cummings (2021) employed HTTP referrer data from six journal publishers and found that 10% of all referral data came directly from the discovery tool, with an average of 10.04%, ranging between 3.22% and 17.83%. Evans and Schonfeld (2020) found that only 6% of users in the OhioLINK network began their discovery process at the library. Hayes, Henry, and Shaw (2021) surveyed over 4,000 librarians and users and discovered that 79% of faculty and 74% of students started their discovery process outside of the library. In terms of starting points for electronic resource discovery, their analysis showed that Google Scholar (24–30%) was the primary starting point, followed by search engines (18–28%) and the library (21–28%). On the other hand, the study by Greenberg and Bar-Ilan (2017) showed that the number of full-text downloads based on COUNTER requests is 3.1 times higher than the downloads requested through the discovery tool, as documented by the library's OpenURL link resolver service.

Our study found that the library's discovery tool was the starting point for accessing ebooks, with an average of 21.74%. This data was calculated using Unique_Title_Investigations and the number of Clicked Requests for each platform. The results revealed significant variations among the individual platforms: EBSCOhost (86.90%), Cambridge Core (6.54%), and ScienceDirect (4.64%). It is worth noting that the primary starting point to access EBSCOhost is the library's discovery tool. This is because this content provider restricts access to its electronic resources via Google and users cannot interact with the platform unless they are authenticated by the university.

### 5.2. Limitations of Library Metrics

Despite COUNTER usage reports offering accurate data on users' interactions with ebook titles, there are certain aspects that are not reflected in these reports. These limitations should be considered when interpreting library metrics.

- COUNTER reports do not provide information on where users downloaded an ebook, whether it was from the library's discovery tool, the vendor platform, or Google.
- COUNTER reports only include the usage of electronic resources if users are connected to the university's authentication system or are on campus. This means that clicks made from the discovery tool by users who are not within the university's IP range cannot be attributed to the university. As a result, investigations made by unauthenticated users from the library's discovery tool are not reflected in the COUNTER reports.

Link resolver statistics only capture a small fraction of total ebook usage, accounting for an estimated average of 21.74% of ebooks accessed through the library's discovery tool in this study. Additionally, link resolver reports have other inherent limitations, including the following:

- Link resolver reports do not differentiate between clicked requests that initiate the process of viewing or downloading an item's full text and those processes that are only investigations.
- Link resolver reports do not distinguish whether a user was able to successfully retrieve the desired article, download the full text, or simply investigate a content item.
- Link resolver reports do not offer information regarding whether the user is connected to the university's authentication system or not.
- Link resolver reports do not provide information about whether an ebook is licensed, purchased, available as Gold Open Access, or simply available via Open Access.
- The OpenURL metadata "genre," which describes the unit of content accessed by a user (rft.genre=bookitem, rft.genre=book), is not always included in the OpenURL syntax description. Furthermore, the use of this metadata can often be problematic when the term *book* is employed as a broad category and is applied to other kinds of materials, including videos (Bulock, 2019).

### 5.3. Practical Applications of Link Resolver Data

The findings of this study have potential applications for tracking and evaluating ebook usage, particularly for titles without COUNTER data. Link resolver statistics may be used to extract information from titles lacking COUNTER data, which are accessible through the library's discovery tool, including Open Access titles. The positive moderate correlation found between COUNTER and link resolver could help identify the most heavily used titles and serve as a supplement when vendors offer data that may be missing, incomplete, or unreliable (De Groote et al., 2013). Additionally, link resolver data can assist in identifying anomalies, discrepancies, or unusual usage patterns of library materials for collection management purposes (Pastva, Shank et al., 2018).

Despite these applications, one important limitation of the link resolver is the small fraction of usage data that it can report. As is recognized, the primary starting point for accessing academic information has shifted from the library to Google Scholar (Greenberg & Bar-Ilan, 2017; Hayes et al., 2021). In the current context, where Google Scholar is a part of the user's workflow, some libraries have integrated their institutional link resolver into Google Scholar to increase the accessibility of electronic resources, including ebooks, and to gain information on electronic resource usage. Although this integration offers potential advantages, some experts hold a different view. Cummings (2021) argued that if libraries become dependent on Google, users are more likely to bypass the library catalog and conduct literature searches directly on the search engine.

In parallel, NISO's Open Discovery Initiative (NISO, 2020) provided guidelines to optimize the utility of discovery tools and assess electronic resources. The initiative defines methods for assessing the content providers' levels of participation in discovery tools, streamlines the process by which content providers work with discovery tool vendors, defines models for "fair" linking from discovery tools to publishers' content, and determines which usage statistics should be collected. Overall, this initiative highlights the importance of collaboration between libraries, content providers, and discovery tool vendors to improve the accessibility, discoverability, and assessment of academic information.

## 6. CONCLUSIONS

This study provides evidence of a moderate Spearman correlation between the COUNTER metrics Unique_Title_Investigations and Unique_Item_Investigations, with the clicked investigations/requests reported by link resolver. Additionally, the small differences found in the correlation coefficients between Investigations and Requests suggest a close relationship between these two metrics.

Regarding the values of the usage types (Investigations and Requests), it is important to note that both metrics are not equivalent. Investigations encompass various types of usage, including full-text access, but Requests reflect only access to full-text content. However, the small differences observed in the correlation coefficients of these metrics may suggest that certain types of Investigations, such as browsing and searching, conducted from the discovery tool, were used as a pathway that might have contributed to accessing full-text content.

The moderate correlation between COUNTER and link resolver suggests that these metrics are complementary, capturing different aspects of ebook usage in different contexts. Utilizing the data from content providers and link resolver statistics in conjunction can offer a more comprehensive view of ebook usage.

COUNTER measures ebook usage generated in different environments, such as content platforms, Google, search engines, bookmarks, and table of contents alerts, as well as through the library's discovery tool when authenticated users access electronic resources. Although COUNTER reports provide detailed information about ebook usage made by authenticated users, they do not provide information about the context in which content items or titles were searched, viewed, or downloaded.

Link resolver provides seamless access to electronic resources and plays an important role by capturing all attempts to access ebooks made from the library's discovery tool. This functionality allows libraries to gather usage data of electronic resources and assess the impact of the discovery tool on the total usage of electronic resources, as reported by COUNTER metrics. However, link resolver has certain limitations. It can only capture a small portion of the total ebook usage and cannot distinguish between investigation activities and downloads. Additionally, its metadata may not accurately identify the section types, and the access types of items. Therefore, it is important to note that although link resolver reports provide valuable usage data, they cannot be used as a substitute for COUNTER reports.

The reviewed studies indicate that the library's discovery tool is not the primary source of access to scholarly information. Instead, Google Scholar has become the first starting point for accessing full-text academic content. According to Greenberg and Bar-Ilan (2017), libraries should analyze users' information retrieval behavior and modify their discovery tool interface accordingly. This could provide valuable insights for reassessing the role of the library's discovery tool as an information retrieval system that integrates searches across the library's resources.

## ACKNOWLEDGMENTS

We are grateful to reviewers for their helpful comments on our work.

## AUTHOR CONTRIBUTIONS

Mercedes Echeverria: Conceptualization, Formal analysis, Methodology, Visualization, Writing—original draft, Writing—review & editing. Yacelli Bustamante: Formal analysis, Methodology, Visualization, Writing—review & editing.

## COMPETING INTERESTS

The authors have no competing interests.

## FUNDING INFORMATION

This research was not funded.

## DATA AVAILABILITY

The data sets analyzed in the present study are available on Figshare: https://doi.org/10.6084/m9.figshare.24529384.v6.